\documentclass[nofootinbib,preprint,showpacs,preprintnumbers,amsmath,amssymb,aps,pre]{revtex4-2}

\usepackage[utf8]{inputenc} 
\usepackage{footnote}
\usepackage{epsfig}
\usepackage{natbib}
\usepackage{amsmath}
\usepackage{times}
\usepackage{epstopdf} 
\usepackage{graphicx}
\usepackage{color}
\usepackage{placeins}

\usepackage{afterpage}

\begin{document}

\title{Energy-based stochastic resetting can avoid noise-enhanced stability} %in chaotic Hamiltonian systems}

\author{Julia Cantisán}
\email[]{julia.cantisan@urjc.es}
\affiliation{Nonlinear Dynamics, Chaos and Complex Systems Group, Departamento de
	F\'{i}sica, Universidad Rey Juan Carlos, Tulip\'{a}n s/n, 28933 M\'{o}stoles, Madrid, Spain}

\author{Alexandre R. Nieto}
\affiliation{Nonlinear Dynamics, Chaos and Complex Systems Group, Departamento de
F\'{i}sica, Universidad Rey Juan Carlos, Tulip\'{a}n s/n, 28933 M\'{o}stoles, Madrid, Spain}

\author{Jes\'{u}s M. Seoane}
\affiliation{Nonlinear Dynamics, Chaos and Complex Systems Group, Departamento de
F\'{i}sica, Universidad Rey Juan Carlos, Tulip\'{a}n s/n, 28933 M\'{o}stoles, Madrid, Spain}

\author{Miguel A.F. Sanju\'{a}n}
\affiliation{Nonlinear Dynamics, Chaos and Complex Systems Group, Departamento de
F\'{i}sica, Universidad Rey Juan Carlos, Tulip\'{a}n s/n, 28933 M\'{o}stoles, Madrid, Spain}

\date{\today}

\begin{abstract}
The theory of stochastic resetting asserts that restarting a stochastic process can expedite its completion. In this paper, we study the escape process of a Brownian particle in an open Hamiltonian system that suffers noise-enhanced stability. This phenomenon implies that under specific noise amplitudes the escape process is delayed. Here, we propose a new protocol for stochastic resetting that can avoid the noise-enhanced stability effect. In our approach, instead of resetting the trajectories at certain time intervals, a trajectory is reset when a predefined energy threshold is reached. The trajectories that delay the escape process are the ones that lower their energy due to the stochastic fluctuations. Our resetting approach leverages this fact and avoids long transients by resetting trajectories before they reach low energy levels. Finally, we show that the chaotic dynamics (i.e., the sensitive dependence on initial conditions) catalyzes the effectiveness of the resetting strategy.

\end{abstract}

%\pacs{05.45.Ac,05.45.Df,05.45.Pq}
\maketitle
\newpage
%%%%%%%%%%%%%%%%%%%%%%%%%%%%%%%%%%%%%%%%%%%%%%%%%%%%%%%%%%%%%%%%%%%%%%%%%%%%%%%%%%%%%%%%%%%%%%%%%%%%%%%%%%%%%%%%%

\section{Introduction} \label{sec:Introduction}

Stochastic resetting is a strategy used in search problems to decrease the search time \cite{Evans2011}. Imagine a simple search problem in which a particle, the searcher, is looking for a target in a plane without any previous information, and there are no boundaries or restrictions in its movement. The easiest strategy for the particle is just to freely diffuse until it finds the target. However, this method is not very efficient as the search time diverges. Stochastic resetting is a much more suited strategy as the search time becomes finite. This technique implies resetting the particle to its initial position at specific time intervals and letting it evolve freely between the resetting events. By resetting, we avoid the trajectories that take too long and are far away from the target, and we give them another chance to complete the search rapidly. 

Search processes in nature are often accompanied by a resetting to an initial state. An example of this is animal foraging \cite{Mercado-Vasquez2018}, where animals repeatedly return to the shelter during the search for resources. Another example where a resetting strategy has also been observed to occur naturally is human visual search \cite{Wolfe2004}. The fact that resetting expedites the search may explain its prevalence in nature. Furthermore, the resetting strategy may be used purposely in different contexts, such as chemical reactions \cite{Reuveni2014} and other queueing processes \cite{Bonomo2022} that work like stochastic processes. The goal is to avoid the metastable states that are causing a delay. In the case of chemical reactions, for instance, the process is reset by unbinding the catalyst. For more applications, see \cite{Evans2020, Gupta2022}. 

Stochastic resetting is a hot topic nowadays in the community and much has been published regarding multiple searchers \cite{Bhat2016}, multiple targets \cite{Bressloff2020}, interacting particle systems \cite{Nagar2023}, or non-diffusive processes \cite{Masoliver2019,Kusmierz2014}, to name a few. The case of diffusion when the searcher can move in a bounded region due to a potential has been less studied, although it is a more realistic situation. Brownian particles subjected to various potentials have been studied, including linear \cite{Pal2015}, logarithmic \cite{Ray2020}, V-shaped \cite{Singh2020} and higher-order potentials \cite{Ahmad2019}. Only recently, stochastic resetting in an escape problem has been studied \cite{Cantisan2021b}. In this case, the search process ends when the particle finds the exit and escapes. The authors considered the Kramers problem, where a particle moving in one dimension has to overcome a potential barrier in order to escape. 

Here, we aim to broaden current knowledge of stochastic resetting in escape processes. For that purpose, we study a two-dimensional open Hamiltonian system, the Hénon-Heiles Hamiltonian. We choose this system as it presents a phenomenon called noise-enhanced stability, which is also referred to as noise-enhanced trapping in the context of chaotic scattering \cite{Altmann2010}. This phenomenon delays the escape process for certain noise intensities. We explore if stochastic resetting can counterbalance noise-enhanced stability and expedite the escape process. This interplay was previously explored in \cite{Capaa2022}, where they considered time-based resetting in a one-dimensional system. 

To expedite an escape process from periodic potentials may be useful in many physical situations. We would like to mention Josephson junctions, that can be used for detection of axion-like particles \cite{Pankratov2022} or logical devices \cite{Gordeeva2006}. In that case, the dynamics can be modelled by a particle moving in the washboard potential and the escape of the particle implies a switch between the superconductive and the resistive states. The noise can delay the escape process by the same noise-enhanced stability phenomenon that was mentioned, producing a noise-delayed switching \cite{Fedorov2008}. In this context, stochastic resetting may increase performance by reducing the switching time.

Furthermore, we propose a new protocol for resetting that we call energy-based resetting. Previous studies have considered resetting at fixed intervals of time, or intervals drawn from a certain distribution (typically a Poisson distribution). Also, a protocol with space-dependent resetting rate has been explored \cite{Evans2011a, Roldan2017}. Here, we show that tracking the energy and resetting when certain thresholds are reached is a strategy capable of expediting the escape process and even suppressing the noise-enhanced stability effect.

This paper is organized as following. In Section \ref{sec:2}, we describe the Hénon-Heiles system and the noise-enhanced stability phenomenon. In Sections \ref{sec:3} and \ref{sec:4}, we implement the strategy of stochastic resetting based on time and based on energy, respectively. Later, in Section \ref{sec:5}, we compare both methods. Finally, in Section \ref{sec:conclusions}, we present our concluding remarks. 

\section{Noise-enhanced stability in the Hénon-Heiles}\label{sec:2}

To study the counterbalance of stochastic resetting and the noise-enhanced stability effect, we use as a model the paradigmatic Hénon-Heiles Hamiltonian \cite{Henon1964}. Ever since its emergence in scientific literature as a model of star motion around a galactic center, this system has garnered extensive attention as a prototype of Hamiltonian chaos. The Hamiltonian is given by 
\begin{equation}
	{\cal{H}}(x,y,\dot{x},\dot{y})=\frac{1}{2}(\dot{x}^2+{\dot{y}}^2)+\frac{1}{2}(x^2+y^2)+x^2y-\frac{1}{3}y^3,
\end{equation}
where $x$ and $y$ are the spatial coordinates, and $\dot{x}$ and $\dot{y}$ are the components of the velocity.

\begin{figure}[h]
	\centering
	\includegraphics[clip,height=7.5cm,trim=0cm 0cm 0cm 0cm]{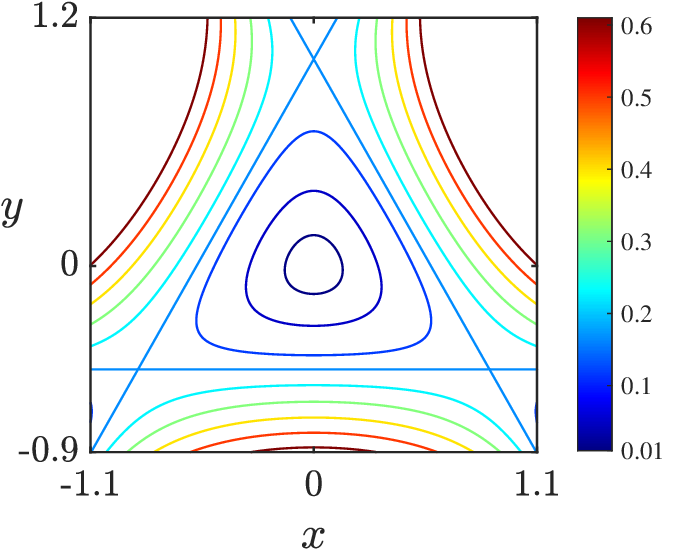}			
	\caption{Isopotential curves of the Hénon-Heiles system for various values of the potential $V(x,y)=\frac{1}{2}(x^2+y^2)+x^2y-\frac{1}{3}y^3$. The color of the curves indicates the different values of the potential, which are marked in the color bar. For energy values below the escape energy $E_e=1/6$, the isopotential curves are closed, while the potential exhibits three symmetrical exits for energy values above $E_e$. Particles can escape through these exits towards $\pm\infty$.}
	\label{Fig0}
\end{figure}

Since the Hamiltonian has no explicit time dependence, the energy is a conserved quantity (i.e., ${\cal{H}}(x,y,\dot{x},\dot{y})=E$). Beyond the critical energy level $E_e=1/6$, which is often referred to as the ``escape energy", the potential has three symmetrical exits separated by an angle of $2\pi/3$ radians, as illustrated in Fig.~\ref{Fig0}. Therefore, in this situation particles can escape towards $\pm\infty$. Conversely, below the escape energy the particles' motion remains bounded. 

\newpage
In order to account for the presence of noise in the system, we include the influence of a random force, which we simplify as an uncorrelated additive Gaussian white noise. Under these considerations, the equations of motion can be expressed as 
\begin{equation} \label{eq_motion}
	\begin{aligned}
		\ddot{x} & = -x - 2xy +\sqrt{2\xi}\eta_x(t), \\
		\ddot{y} &= -y - x^2 + y^2+\sqrt{2\xi}\eta_y(t),
	\end{aligned}
\end{equation}
where $\eta_x(t)$ and $\eta_y(t)$ are Gaussian white noise processes $X_t\sim{\cal{N}}(0,2\xi)$ of amplitude $\xi$ and autocorrelation function $\langle\eta(t')\eta(t)\rangle=\sqrt{2\xi}\delta(t'-t)$.

We have solved numerically the system of stochastic differential equations (\ref{eq_motion}) using a stochastic second-order Heun method \cite{Kloeden1992}. The robustness of this numerical scheme in the stochastic Hénon-Heiles system has been tested in Ref.~\cite{Nieto2021}.

The phenomenon of noise-enhanced stability occurs in the Hénon-Heiles system for certain noise conditions. This phenomenon implies that the average escape time suffers a peak for a specific noise amplitude. To explore this, we launch trajectories from $(x,y)=(0,0)$ using a random shooting angle $\theta\in[0,2\pi/3]$, which is measured counterclockwise with respect to the $y$ axis. Note that, given the symmetry of the system, it is not necessary to consider the complete circumference. Without loss of generality, henceforth we use an initial energy value of $E_0=0.25$. Due to the effects of noise, the energy value exhibits fluctuations, so $E_0$ only defines the initial conditions for the shooting. Once a shooting angle is defined, the components of the initial velocity are given by $(\dot{x},\dot{y})=\sqrt{2E_0}(\sin\theta,\cos\theta)$.

\begin{figure}[h]
	\centering
	\includegraphics[clip,height=7.5cm,trim=0cm 0cm 0cm 0cm]{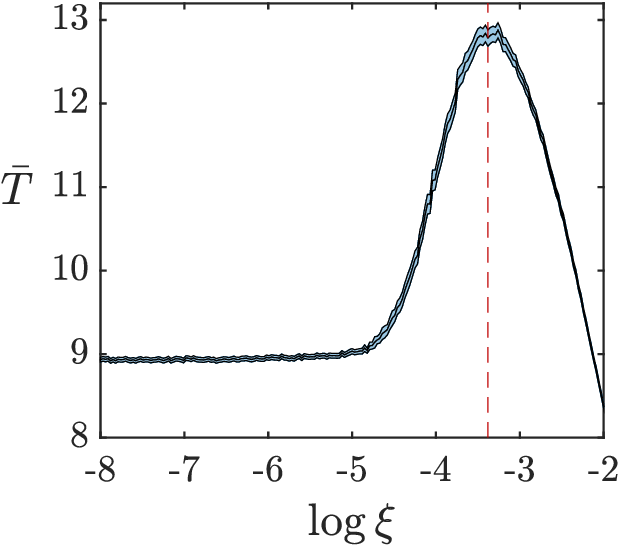}			
	\caption{Average escape time for $200$ different noise amplitudes. The escape time of $10^6$ initial conditions is averaged for each value of $\xi$. The blue shaded region represents a $99.7\%$ confidence interval and the peak, emphasized by a red dashed line, corresponds to the noise-enhanced stability phenomenon.}
	\label{Fig1}
\end{figure}

We represent the average escape time $\bar{T}$ for different noise levels $\xi$ in Fig.~\ref{Fig1}. The blue shaded region around $\bar{T}$ corresponds to a $99.7\%$ confidence interval. The peak corresponding to the noise-enhanced stability phenomenon, indicated by a red dashed line in Fig.~\ref{Fig1}, can be clearly distinguished in our system. In particular, the noise amplitude that enhances the stability corresponds to the value $\xi_s=10^{-3.38}$. This is precisely the effect that we try to counterbalance by applying the resetting, so henceforth we fix $\xi=\xi_s$. For this noise amplitude, the escape time averaged over all the initial angles is $\bar{T}=12.97$.

\begin{figure}[h]
	\centering
	\includegraphics[clip,height=6.8cm,trim=0cm 0cm 0cm 0cm]{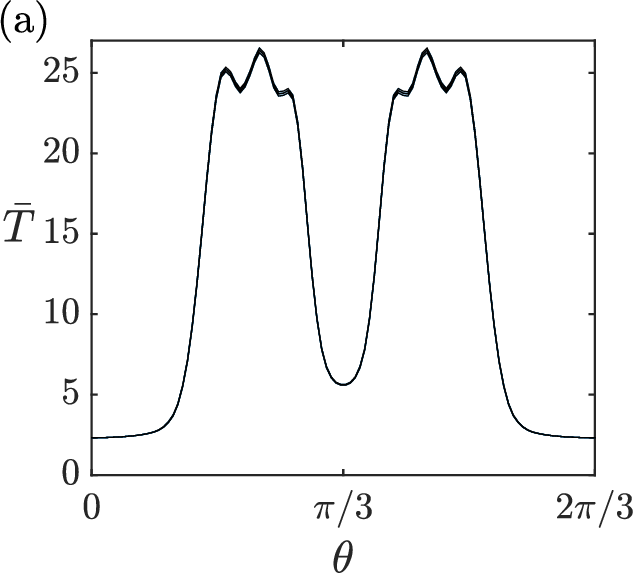}	
	\includegraphics[clip,height=6.8cm,trim=0cm 0cm 0cm 0cm]{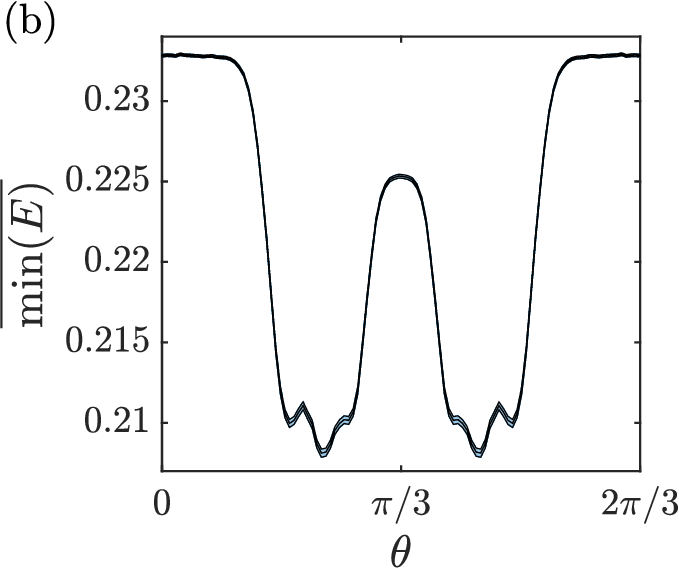}			
	\caption{(a) Average escape times and (b) average minimum energy during the escape path for $100$ different values of $\theta$. In both panels, $10^6$ initial conditions are computed for each angle.}
	\label{Fig2}
\end{figure}

It is clear that for a certain noise level the escape time depends on the launching angle. To observe this, we represent the average escape times for different values of $\theta$ in Fig.~\ref{Fig2}(a). As one can see, trajectories initially pointing directly to the exits take less time to escape than others. The same occurs for trajectories launched with angles close to $\pi/3$, which are initially perpendicular to the potential. It is also remarkable that the standard deviation increases for the angles for which $\bar{T}$ increases. This can be observed in the blue shaded region representing the confidence intervals, which is widened for those angles. 

Furthermore, in Fig.~\ref{Fig2}(b) we represent the average minimum energy that the trajectories reach before escaping. Interestingly, it presents mirror symmetry compared to Fig.~\ref{Fig2}(a). This implies that fast escaping particles do not have the possibility to lower their energy. On the contrary, the particles that remain in the scattering region during long transients can wander in the potential well at low energy states.

\begin{figure}[h]
	\centering
	\includegraphics[clip,height=7.5cm,trim=0cm 0cm 0cm 0cm]{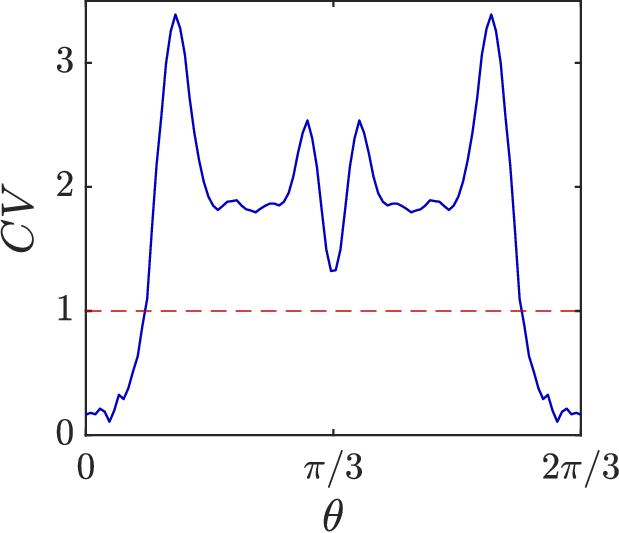}			
	\caption{Coefficient of variation, $CV=\sigma(T)/\bar{T}$, for $100$ values of $\theta$. For each value of $\theta$, the average escape time and the corresponding standard deviation have been estimated from $10^6$ realizations of the process.} %Resetting to angles with $CV>1$ expedite the escape process.
	\label{FigCV}
\end{figure}

Since our purpose is to study if stochastic resetting is a valid strategy to avoid the noise-enhanced stability effect, we shall compute the coefficient of variation ($CV$) beforehand. This metric measures the ratio of the standard deviation, $\sigma$, and the mean value of the escape times, i.e., $CV=\sigma(T)/\bar{T}$. In \cite{Pal2017}, the authors show that whenever $CV>1$, stochastic resetting can expedite the search process. The reasoning behind this is that $CV>1$ implies a wide distribution of search times, with certain trajectories taking a large time to complete the process. These are the trajectories that delay the whole process and that are given a second chance by resetting. Taking that into consideration, we explore which angles are more beneficial for resetting in Fig.~\ref{FigCV}.

We see that $CV>1$ for all angles, except the ones close to the exits. The fundamental reason for this is that initial conditions launched directly towards the exit generate non-chaotic trajectories, that are not worth resetting. In Hamiltonian systems, chaos occurs due to the existence of a non-attracting chaotic set (a chaotic saddle) in phase space. The pattern that can be observed for $CV>1$ is explained by the structure of the stable manifold of the chaotic saddle in the noiseless system. This stable manifold is formed by the same set of points as the basin boundary. This implies that orbits that start close to the basin boundary, are the ones that take longer times to escape as they follow the stable manifold to the chaotic saddle, spend long times in its vicinity and finally escape following the unstable manifold. If now we consider the noise, an orbit starting close to the stable manifold (i.e., the basin boundary) can jump to different basins and consequently different realizations of the process can have drastically different escape times. Ultimately, this leads to $CV>1$. In this sense, it is worth noting that the chaotic dynamics (i.e., the sensitive dependence on initial conditions) catalyzes the effectiveness of the resetting strategy.

\begin{figure}[h]
	\centering	
 \includegraphics[clip,height=6.8cm,trim=0cm 0cm 0cm 0cm]{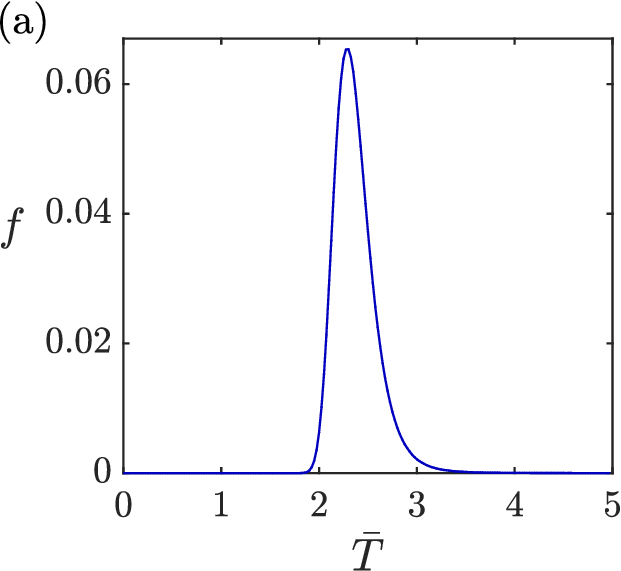}
	\includegraphics[clip,height=6.8cm,trim=0cm 0cm 0cm 0cm]{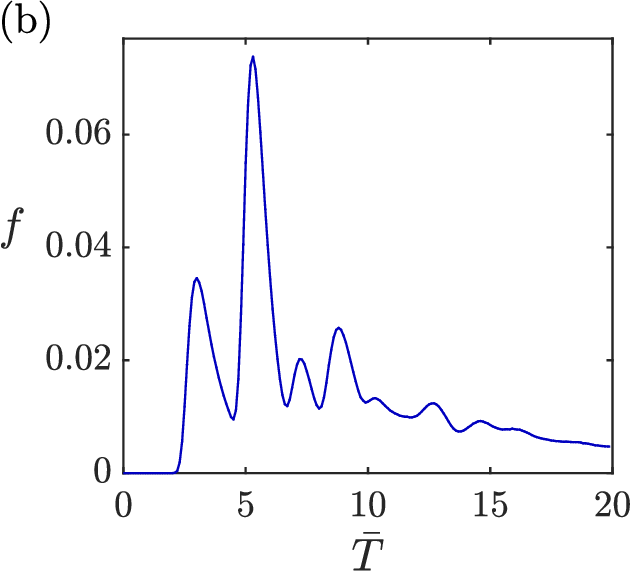}
	\caption{Escape time distribution based on initial conditions with (a) low $CV$ and (b) high $CV$. The escape time is averaged over $10^7$ initial conditions in the intervals: $\theta\in[0,0.24]\cup [1.85,2\pi/3]$ for low $CV$ and $\theta\in[0.37,1.71]$ for high $CV$. The relative frequency $f$ is normalized to unity. The distribution for high $CV$ shows a long tail of escape times, which we aim to avoid by resetting.}
	\label{Fig:distribution}
\end{figure}

All of the above results can be better understood by looking at the distribution of escape times for two characteristic ranges of angles: one with high $CV$ ($\theta\in[0.37,1.71]$) and one with low $CV$ ($\theta\in[0,0.24]\cup [1.85,2\pi/3]$). The escape time is averaged for $10^{7}$ initial conditions in those intervals and represented in the form of histograms. In Fig.~\ref{Fig:distribution}(a), the relative frequency $f$ (normalized to unity) of the escape time for the trajectories with low $CV$ is depicted. The distribution is narrow, trajectories escape fast and are not worth resetting. Meanwhile, in Fig.~\ref{Fig:distribution}(b) we can see trajectories from a range of angles with a high $CV$. In this case, the distribution shows a long tail and trajectories are worth resetting. 

\FloatBarrier
\section{Resetting based on time} \label{sec:3}

\begin{figure}[h]
	\centering
	\includegraphics[clip,height=7.5cm,trim=0cm 0cm 0cm 0cm]{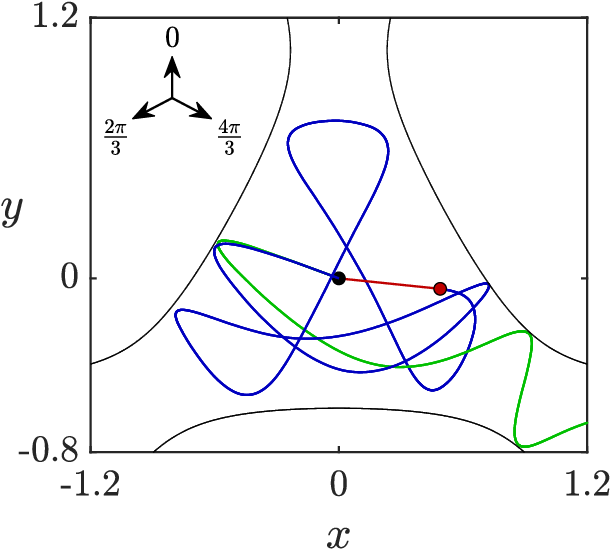}			
	\caption{Stochastic resetting for a trajectory in the Hénon-Heiles Hamiltonian system. The angles $\theta$ for the exits are marked in the upper left corner. We show a trajectory (blue curve) that starts at the origin (black dot). The position of the particle at $t_r$ is marked with a red dot, while the resetting process is represented with a red segment. After resetting, the new trajectory (green curve) escapes in a time $t<t_r$.}
	\label{FigSRvisualization}
\end{figure}

We start applying resetting to our system at deterministic time intervals. This technique has also been called sharp restart \cite{Evans2023}. We fix the resetting position to the given initial condition $(x,y,\dot{x},\dot{y})=(0,0,\sqrt{2E_0}\sin\theta, \sqrt{2E_0}\cos\theta)$, where $E_0=0.25$. This implies that a particle launched with a certain angle is left to evolve freely in the potential until a certain time $t_{r}$, which is the resetting time interval. Then the particle is reset to the initial position and launched again with the same velocity and angle as in the beginning. This process is repeated as many times as needed until the particle escapes. The whole process is represented in Fig.~\ref{FigSRvisualization}, where the black dot denotes the initial position and the black curves account for the potential. The trajectory in blue wanders until the resetting time interval $t_{r}$. At that time, the position is marked by a red dot. Then the trajectory is reset (see red segment). The process is then restarted and now the particle escapes through the lower right exit after a short time (see green trajectory). We recall that both trajectories are not the same due to the noise. Furthermore, their path is completely different because they are chaotic.

\newpage
In the remainder of this section, we show that time-based stochastic resetting can expedite the escape and we find the optimal resetting time interval $t_r^*$ to that end. As we have seen, not all trajectories are worth resetting. In particular, for initial conditions with $CV<1$ stochastic resetting is detrimental. Therefore, the resetting strategy should be focused on initial conditions with $CV>1$ (i.e., the chaotic ones). This already points out that $t_{r}^{*}$ should be larger than the escape time of the initial conditions with $CV<1$ (recall Fig.~\ref{Fig:distribution}(a)).

\begin{figure}[h]
	\centering
    \includegraphics[clip,height=6.8cm,trim=0cm 0cm 0cm 0cm]{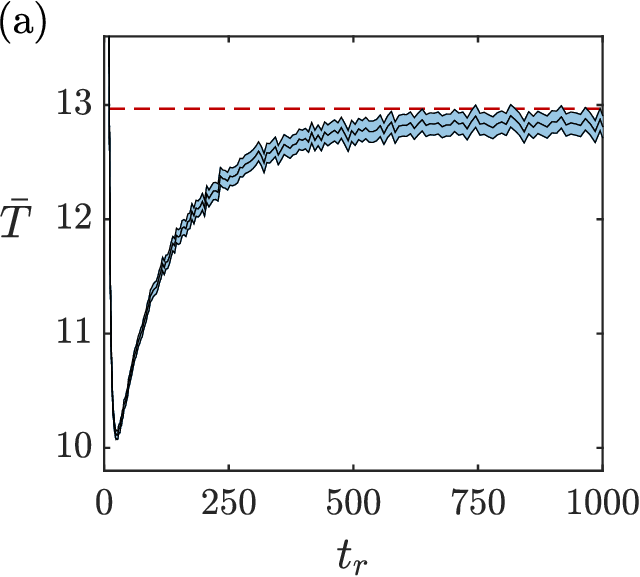}
     \includegraphics[clip,height=6.8cm,trim=0cm 0cm 0cm 0cm]{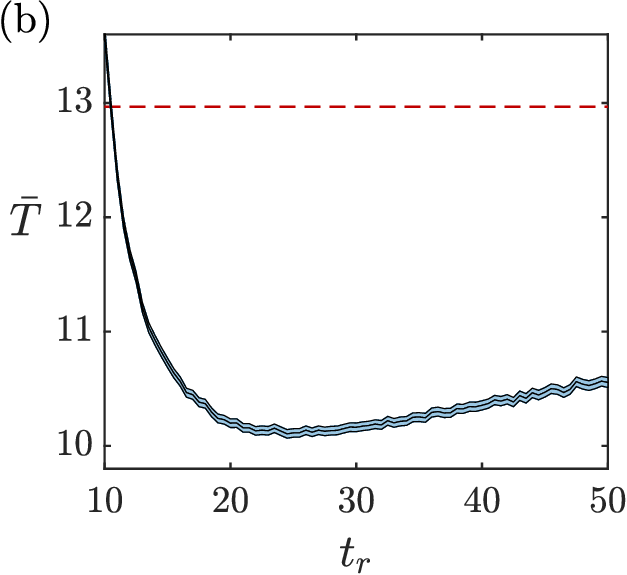}
	\caption{Escape time for $200$ values of the resetting interval $t_r$. For each value of $t_r$, $\bar{T}$ has been calculated by averaging the escape time of $10^6$ random initial conditions. In (b) we show a zoom around the optimum $t_r$. The horizontal red dashed line marks the average escape time of the system without stochastic resetting ($\bar{T}=12.97$). }
	\label{Fig3}
\end{figure}

In Fig.~\ref{Fig3}, we show how the escape process is affected by different values of the resetting interval $t_{r}$. Panel (b) is a magnification of panel (a), focused on the interval of values of $t_r$ where stochastic resetting is more effective. In both panels the horizontal dashed red line marks the average escape time of the system without resetting ($\bar{T}=12.97$). For $t_{r}<10.5$ stochastic resetting hinders the escape process. We have already mentioned that $t_{r}^{*}$ should be larger than the average escape time for trajectories with $CV<1$, as these trajectories should not be reset. For the initial conditions with $CV>1$, the optimum strategy is the one that does not reset the trajectories that escape fast, but only the ones that belong to the tail of escape times. Stochastic resetting starts to be effective for $t_{r}>10.5$. If we compare this value with Fig.~\ref{Fig:distribution}(b), it implies resetting after the big waves of escape have passed. For example, the probability that an initial condition with initial angle $\theta=0.7$ (maximum in Fig.~\ref{Fig2}(a)) escapes before $t=5$ is $2\cdot 10^{-6}$. The probability that it escapes before $t=3$ is $0$ after launching $10^6$ simulations. This implies that resetting has to occur for larger values of $t_{r}$. Otherwise, we would be resetting again and again without letting any trajectory to escape. On the other hand, for very large values of $t_{r}$, stochastic resetting is not beneficial either, as it is equivalent to not applying any protocol. As a matter of fact, from Fig.~\ref{Fig3}(a) it is clear that for large values of $t_r$ the escape time converges to the natural value of the system without resetting.

The optimum resetting time interval is $t_r^*\approx 25$. For this value we find $\bar{T}=10.11$, which is a $22\%$ less than the average escape time of the system without stochastic resetting. Furthermore, there is a wide range of values above $t_r^*= 25$ that also bring good results, thus the chosen resetting interval does not need to be extremely accurate. 

An advantage of this method is that the fraction of reset trajectories is low (only $9.3\%$). This is convenient if we assume that the reset implies a certain cost. A simplification we have taken in our model is that the resetting is instantaneous. Of course, every resetting event comes with a cost and requires a certain time. This has been implemented by including refractory periods or time penalties proportional to the distance to the resetting position. Energetic and thermodynamics costs have also been studied. For a review, see \cite{Sunil2023}.

\section{Resetting based on energy} \label{sec:4}

In this section, we introduce a novel approach for resetting that is based on the energy of the particle. In this case, the particle launched with a certain angle is left to evolve freely in the potential, until a certain energy threshold $E_{r}$ is reached. Then the trajectory is reset to the initial position and launched with the same velocity and angle as in the beginning. 

The interest of this approach is the following. Noise-enhanced stability occurs due to a small amount of trajectories that decrease their energy to very low values and remain during long transients in the scattering region. This drop in the energy allows these `unusual' trajectories to wander in Kolmogorov-Arnold-Moser (KAM) regions or even to reach energy levels for which the isopotential curves are closed \cite{Nieto2021a}. Thus, the more natural parameter to track seems to be the energy. This is not only true for our specific system, but for all systems in which the process gets stuck in metastable states. 

 \begin{figure}[h]
	\centering
	\includegraphics[clip,height=6.8cm,trim=0cm 0cm 0cm 0cm]{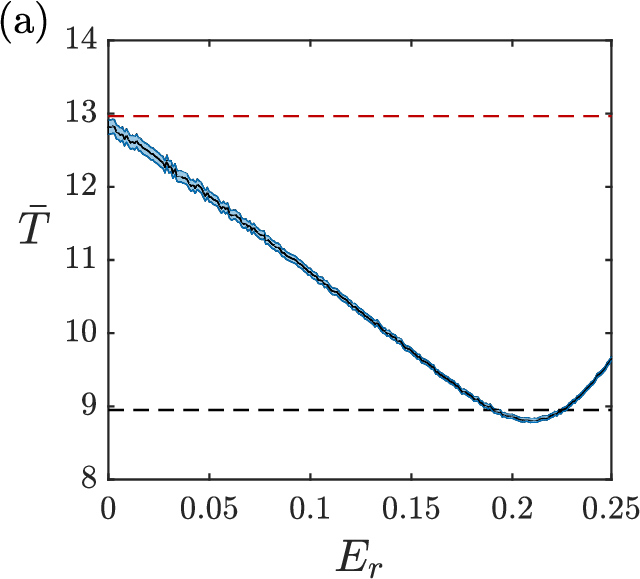}	
	\includegraphics[clip,height=6.8cm,trim=0cm 0cm 0cm 0cm]{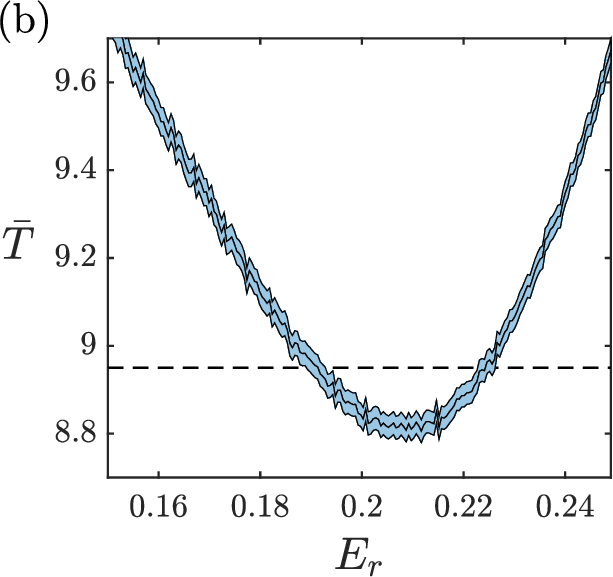}					
	\caption{Escape time averaged over $10^6$ random initial conditions for $250$ different values of the resetting energy $E_r$. In (b) we show a zoom around the optimum value of $E_r$. The horizontal red dashed line is located at the average escape time of the system without stochastic resetting ($\bar{T}=12.97$). The horizontal black dashed line is located at the average escape time of the noiseless system ($\bar{T}=8.95$). It can be seen that resetting based on energy is capable of avoiding the noise-enhanced stability effect and even reducing the escape time under the value of the noiseless system. }
	\label{Fig8}
\end{figure}

Following this protocol, we show how the resetting energy $E_{r}$ affects the escape times in Fig.~\ref{Fig8}(a). For the sake of comparison, the horizontal red dashed line marks the average escape time of the system without resetting. First of all, the initial energy is fixed in all cases to $E_{0}=0.25$ and we can see that if the energy drops below this value, resetting is always beneficial in the sense that it accelerates the escape process. However, the behaviour is non-monotonical and the optimum value of $E_r$ can be found at $E_r^*=0.208$.

The origin of the minimum in the escape times for resetting with $E_r^*$ can be explained as follows. On the one hand, if we set the threshold at too low energies, the escape time increases. This is because trajectories do not reach so low energy values in short times. Thus, setting a low $E_{r}$ would imply resetting after large periods of time. On top of that, only a few trajectories reach energy values close to zero, so we would be resetting only these ones and not others that also take large times to escape. On the other hand, setting $E_r$ too close to the initial energy level results in resetting a high number of particles, including those that were about to escape fast after experiencing an initial energy drop. The action of both effects leads to the nonmonotonic behavior and the presence of an optimum value for $E_r$. In this sense, the method predicts the trajectories that will take longer times and resets them beforehand.

   \begin{figure}[h]
 	\centering
 	\includegraphics[clip,height=6.4cm,trim=0cm 0cm 0cm 0cm]{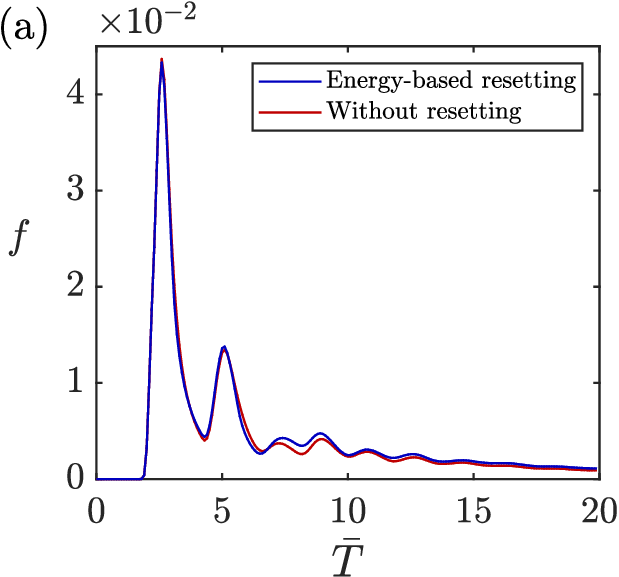}
 	\includegraphics[clip,height=6.4cm,trim=0cm 0cm 0cm 0cm]{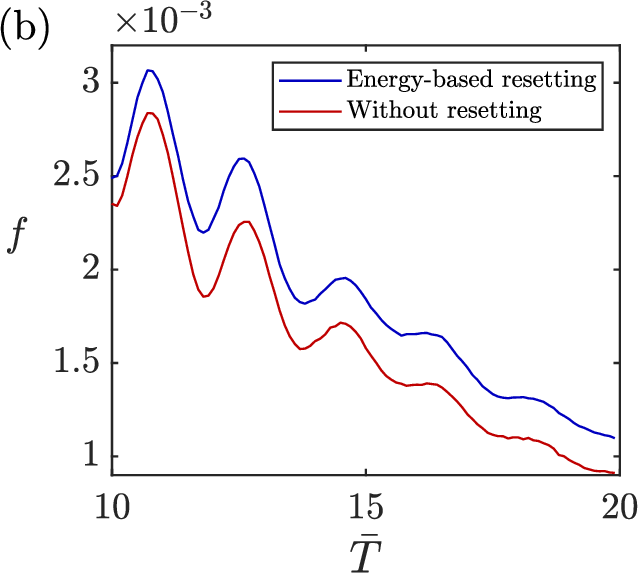}		
 	\includegraphics[clip,height=6.4cm,trim=0cm 0cm 0cm 0cm]{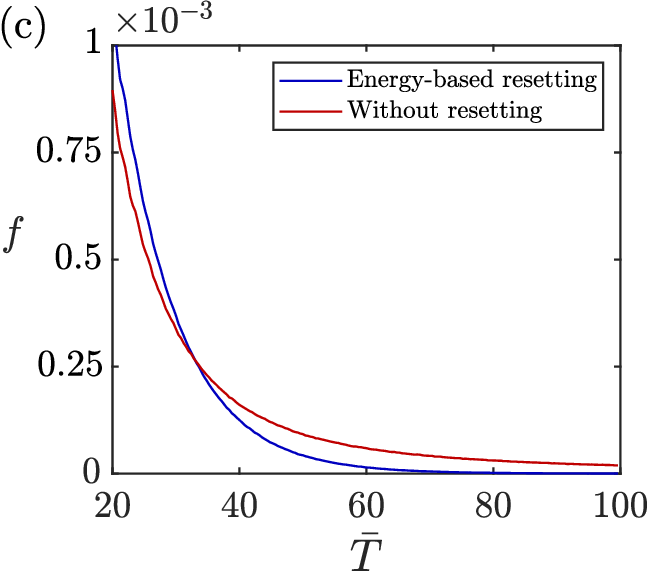}	
 	\includegraphics[clip,height=6.4cm,trim=0cm 0cm 0cm 0cm]{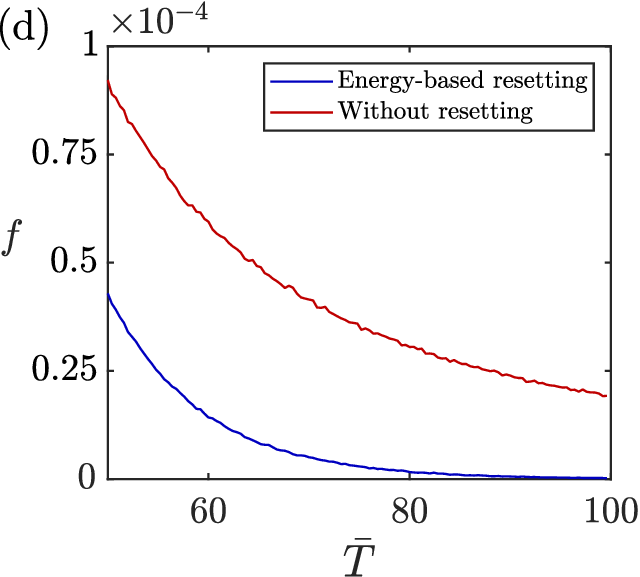}		
 	\caption{Escape time distribution averaged over $10^8$ initial conditions. The relative frequency $f$ is normalized to unity. The red curve is calculated for the system without resetting, while the blue curve denotes the system with stochastic resetting with $E_r=E_r^*=0.208$. Panels (b) and (d) are zoom-ins of panels (a) and (b), respectively. It can be seen that the resetting is capable of reducing the long tail of escape times.}
 	\label{Fig9}
 \end{figure} 

In Fig.~\ref{Fig8}(b), we can see a zoom of Fig.~\ref{Fig8}(a) to observe that the value of $\bar{T}$ drops slightly below the horizontal dashed black line, which denotes the average escape time of the noiseless system ($\bar{T}=8.95$). Thus, we can conclude that energy-based resetting is able to suppress noise-enhanced stability. For this value, the average escape is $\bar{T}=8.81$, which is a $32\%$ less than than the average escape time of the system without stochastic resetting. Also, the maximum escape time is $\mbox{max}(T)=163$, while without stochastic resetting $\mbox{max}(T)>10000$ and after $T=163$ still a $0.8\%$ of the trajectories did not escape. These trajectories have $\bar{T}=295$ and constitute the $20\%$ of the average escape time. These are the type of events that stochastic resetting is good at avoiding. We recall that we have fixed the noise level $\xi$ to the value that maximizes the noise-enhanced effect (see Fig.~\ref{Fig1}), so we have shown that the method is able to suppress noise-enhanced stability even in the worst scenario. Needless to say, the method is also efficient for other values of $\xi$. 

In Fig.~\ref{Fig9}, we depict the escape time distribution for the case without resetting (red curve) and with energy-based resetting using $E_{r}^{*}$ (blue curve). Figures \ref{Fig9}(b) and (d) are zooms of (a) and (c), respectively. The method's primary advantage lies in the reduction of escape times' tail. 

\begin{figure}[h!]
	\centering
	\includegraphics[clip,height=6.9cm,trim=0cm 0cm 0cm 0cm]{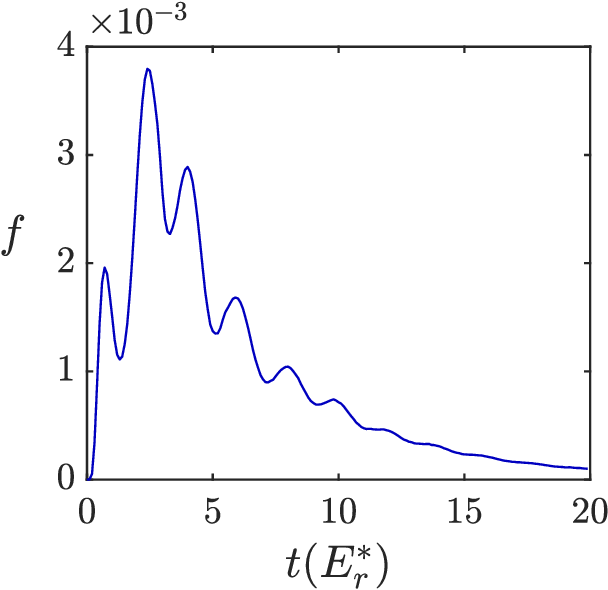}	
	\caption{Distribution of times at which the energy-based resetting with $E_{r}^{*}$ is applied. The results of this figure are based on $10^{8}$ initial conditions and the relative frequency $f$ is normalized to unity. It can be seen that this protocol takes place mainly at early stages of the process.}
	\label{Fig10}
\end{figure}

Another interesting result is that $19.5\%$ of the trajectories are reset using $E_r^*$. Thus, more trajectories are reset using this method than using the sharp restart. Furthermore, the position of the particles when they are reset following the energy-based protocol is approximately uniform, so there is no relation between the position and the resetting in this case. Regarding the times at which the energy-based resetting with $E_{r}^{*}$ is applied $(t(E_{r}^{*}))$, we have computed the distribution and the results can be seen in Fig.~\ref{Fig10}.

As previously mentioned, the noise-enhanced stability is caused by a small amount of trajectories that decrease their energy to very low values and remain during long transients in the scattering region. For this phenomenon to occur, trajectories must avoid the escape by decreasing their energy fast. If not, fluctuations would probably kick the particle out of the well. Therefore, the particles that enhance the stability reach low energy values in short times. Energy-based resetting identifies these trajectories by detecting a fast energy drop. As we can see in Fig.~\ref{Fig10}, most of the trajectories are reset after very short times, meaning that, on average, it is not necessary to let a trajectory evolve to know that its escape time will be large. This is one of the key advantages of energy-based resetting in comparison with time-based resetting. 

\FloatBarrier
\section{Comparison between methods} \label{sec:5}

In this section, we will make a comparison between time and energy-based resetting to show the advantages and downsides of both methods. Furthermore, the results from the preceding two sections have been calculated after averaging different initial conditions using a random shooting angle $\theta \in [0, 2\pi/3]$. Here, we explore the effects of both methods depending on the launching angle.

\begin{figure}[hbt!]
	\centering
 \includegraphics[clip,height=6.9cm,trim=0cm 0cm 0cm 0cm]{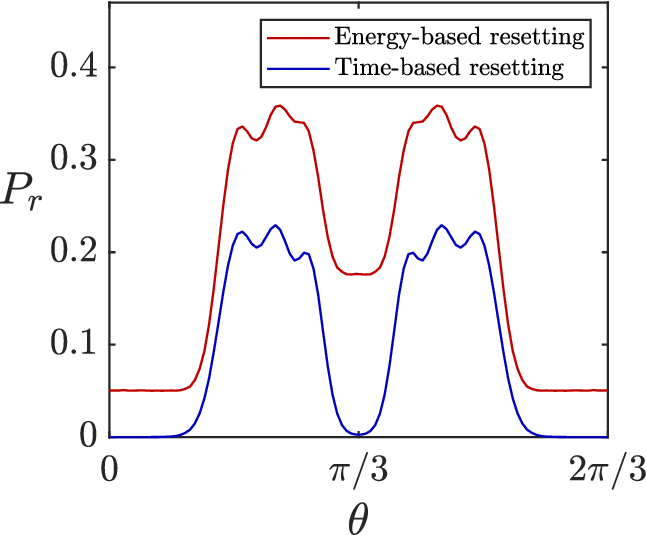}		
	\caption{Resetting probability for $100$ different values of the launching angle. In blue, for the time-based resetting protocol with $t_r^{*}=25$ and in red for the energy-based resetting protocol with $E_r^{*}=0.208$. For each initial condition, we perform $10^{6}$ realizations of the process in order to obtain the resetting probability.}
	\label{Fig11}
\end{figure}

We define the resetting probability $P_r$ as the probability that an initial condition is reset using a particular value of $t_r$ or $E_r$. In the case of time-based resetting, we fix the resetting time to $t_r^*$ and for $100$ values of the angle in the interval $\theta\in[0,2\pi/3]$ we compute $10^6$ realizations of the process. From these data, we calculate the fraction of trajectories that have not escaped after $t=t_r^*$. Equivalently, for energy-based resetting we fix the resetting energy to $E_r^*$ and we calculate the fraction of trajectories that have reached $E=E_r^*$ before escaping. With this procedure, we are calculating the fraction of initial conditions that are reset at least once using the time or energy-based protocols. Therefore, the resetting probability is a measure of the cost of each method.
 
The dependence of $P_{r}$ with the angle is depicted in Fig.~\ref{Fig11}. For time-based resetting (blue curve), stochastic resetting is never applied for angles with $CV<1$. This is simply because the non-chaotic trajectories generated by these initial conditions always escape before $t=t_r^*$. However, the energy-based protocol (red curve) is applied even in those cases, although it is unnecessary. 

\begin{figure}[h]
	\centering
 \includegraphics[clip,height=6.9cm,trim=0cm 0cm 0cm 0cm]{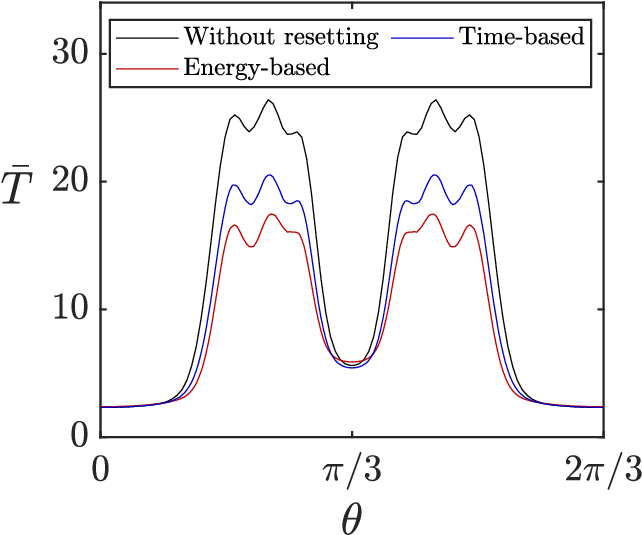}		
	\caption{Escape time for $100$ different values of the launching angle. In blue, for the time-based resetting protocol with $t_r^{*}=25$ and in red for the energy-based resetting protocol with $E_r^{*}=0.208$. For comparison, we show in black the escape times without applying resetting. For each initial conditions, we perform $10^{6}$ realizations in order to obtain $\bar{T}$.}
	\label{Fig12}
\end{figure}

To study how effective the method is depending on the angle, we compute the escape times for different angles in Fig.~\ref{Fig12}. The blue (red) curve corresponds to time-based (energy-based) resetting. The black curve indicates the escape time of the system without stochastic resetting. It is clear that energy-based resetting is, overall, more effective than time-based resetting. In particular, the average escape time of the trajectories close to the basin boundary is reduced much more significantly than using time-based resetting. On the contrary, for initial conditions far from the basin boundary this method is ineffective and can be even counterproductive. This is the case for $\theta=\pi/3$ and for $\theta$ close to $0$ and $2\pi/3$. This suggests that a combined method (i.e., using energy-based resetting in conjunction with time-based resetting) could be beneficial. Energy-based stochastic resetting resets the trajectories in very short times (recall Fig.~\ref{Fig10}), so for initial conditions far from the boundary (non-chaotic ones) the method also resets trajectories that were going to escape very fast, making the method inefficient for such initial conditions. By forcing the reset to occur only if the time is greater than some threshold ($t_r\approx 3$) or if $CV>1$, we could retain the advantages of the method while reducing its downsides.

\begin{figure}[h]
	\centering
 	\includegraphics[clip,height=7.5cm,trim=0cm 0cm 0cm 0cm]{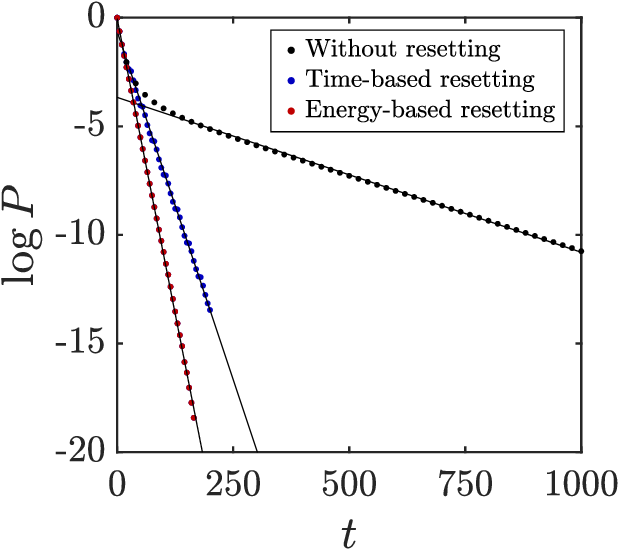}	
	\caption{Survival probability decay with time. Blue dots correspond to stochastic resetting based on time with $t_r=25$ [$\alpha=0.0641$ (r=0.9991)], red dots for stochastic resetting based on energy with $E_r=0.208$ [$\alpha=0.1087$ (r=0.9997)], and black dots for the system without stochastic resetting [$\alpha=0.0071$ (r=0.9993)]. The results for each data set are calculated after shooting $10^8$ initial conditions. }
	\label{Fig13}
\end{figure}

Another consequence of resetting is a reduction on the survival probability, that is, the probability that a certain trajectory has not escaped until time $t$. This is a standard measure in scattering problems in open Hamiltonian systems and for fully chaotic systems it roughly follows an exponential decay law
\begin{equation}
	P(t)\propto e^{-\alpha t},
\end{equation}
where $1/\alpha$ is commonly referred to as the characteristic time. In Fig.~\ref{Fig13}, we see in logarithmic scale how this probability changes with time for the case without resetting (black), with time-based resetting (blue) and with energy-based resetting (red). In this figure, we see that $\alpha$ (the slope of the straight lines) is larger for energy-based stochastic resetting. Although both methods significantly increase the escape rate, energy-based resetting is a better protocol to expedite the escape process.

\FloatBarrier
\section{Conclusions and discussion}\label{sec:conclusions}

Stochastic resetting has been proven to be a very effective protocol to reduce search time in search processes, and recently the same has been found for escape processes too. This strategy consists on restarting a stochastic process in order to avoid the system getting stuck in an undesired state. 

Here, we have studied an open Hamiltonian system that presents noise-enhanced stability. This implies that certain noise levels delay the escape process. For these values, some trajectories present long transients as the noise pushes them close to KAM islands. Our aim has been to explore if stochastic resetting can counterbalance this effect and expedite the escape process. This technique could be applied in many physical contexts involving the escape from periodic potentials, such as to improve the performance of Josephson junctions.

We have explored two different resetting protocols: a time-based protocol (sharp restart) and a novel protocol based on tracking the energy of the system and resetting the process when a certain threshold is reached. We have proven that this protocol is capable of avoiding the noise-enhanced stability phenomenon. Although time-based resetting is not as effective in reducing the escape time, each protocol has its advantages and disadvantages. Resetting is expected to have a cost, either in terms of time or in terms of energy. In this sense, time-based resetting is applied to a smaller number of trajectories, which probably reduces the cost of resetting. Energy-based resetting tends to over-reset, so even for trajectories that have $CV<1$ resetting is sometimes applied. Also, tracking the energy of the process might not be possible in all systems and time-based energy does not imply tracking any measure of the process. 

Another important remark is that we have explored stochastic resetting in a system that presents transient chaos. We conclude that in this situation the effectiveness of the resetting strategy is strongly linked to chaotic dynamics. Trajectories exhibiting sensitive dependence on initial conditions can escape in a wide range of escape times under the effects of noise. This implies high values of $CV$ and thus stochastic resetting is a more interesting strategy to be applied.

 \section*{ACKNOWLEDGMENTS}
This work has been financially supported by MCIN/AEI/10.13039/501100011033 and by “ERDF A way
of making Europe” (Grant No. PID2019-105554GB-I00).

%\bibliography{SRHH}
%

\end{document}